# Consistency Analysis of Sensor Data Distribution

Gianluca Reali and Mauro Femminella
Dipartimento di Ingegneria Elettronica e dell'Informazione, University of Perugia, Perugia, Italy
gianluca.reali@diei.unipg.it, mauro.femminella@diei.unipg.it

*Abstract*— In this paper we analyze the probability of consistency of sensor data distribution systems (SDDS), and determine suitable evaluation models. This problem is typically difficult, since a reliable model taking into account all parameters and processes which affect the system consistency is unavoidably very complex. The simplest candidate approach consists of modeling the state sojourn time, or holding time, as memoryless, and resorting to the well known solutions of Markovian processes. Nevertheless, it may happen that this approach does not fit with some working conditions. In particular, the correct modeling of the SDDS dynamics requires the introduction of a number of parameters, such as the packet transfer time or the packet loss probability, the value of which may determine the suitability of unsuitability of the Markovian model. Candidate alternative solutions include the Erlang phase-type approximation of nearly constant state holding time and a more refined model to account for overlapping events in semi-Markov processes.

## I. INTRODUCTION

The evolution of sensor data distribution systems (SDDS) includes solutions for providing the means for making sensor data available to different recipients. In particular, real time SDDS have to fulfill this task within a timeframe of a few seconds or even milliseconds. Information systems requiring the exchange of real-time data may include tactical networks, traffic monitoring networks, and civil protection networks.

The aim of this paper is to provide suitable analytical models to analyze the consistency of SDDS. This problem has recently become important in SDDS due to their increasing complexity and the need of synchronizing sensor operation in many applications.

In what follows, we assume that the protocol used to communicate with sensor is the Constrained Application Protocol (CoAP) [15], which has been defined by the IETF Constrained RESTful Environments (CoRE) working group. Endpoint interactions may happen asynchronously, according to a request/response model. CoAP implementations may typically play both client and server roles. A CoAP request is issued by a client, typically an application server [3] or a sensor gateway [4], in order to request specific actions to servers, executed by sensors.

The consistency problem is significant in both communication ways. We define an SDDS as *outbound consistent* if a sensor configuration or a data request message, issued by a central repository, is correctly received by all intended recipients. An SDDS is defined as *inbound consistent* if the information stored within a central repository, having the task of collecting it, is piece-wise identical to the relevant information stored within the remote senders.

Our analysis aims to find the *consistency probability*, which is the probability that an SDDS is consistent. To face this problem we first model the consistency condition as a *state* of the analyzed system and then find the probability of such a state. The number of states and the statistics of their sojourn time determine the mathematical model usable to find the consistency probability. The simplest candidate approach consists of modeling the state sojourn time, or holding time, as memoryless, and resorting to the well known solutions of Markovian processes. Nevertheless, it may happen that this approach does not fit with some working conditions. In particular, the correct modeling of the SDDS dynamics requires the introduction of a number of parameters, such as packet transfer time or packet loss probability, the value of which may determine the suitability of unsuitability of the Markovian model. A suitable model accounting for generic state holding time consists of semi-Markov processes. Since we need to represent nearly constant state holding time, we approximate them by using an Erlang phase-type distribution [5][14]. First, we propose a complex model that accounts for overlapping events. Then, we simplify this model in order to reduce its complexity and evaluate its suitability. Results are validated by comparing them with those obtained by determining an analytical solution of the semi-Markov model.

In what follows, we analyze two SDDS case studies depending on some key values, and for both of them we determine the most suitable approach for calculating the consistency probability. Suitability of the analyzed alternative solutions are assessed by considering both reliability and simplicity. In section II we describe the statistical model. In section III, we analyze some case studies. The relevant related works is reported in section IV, and section V includes our concluding remarks.

## II. PROBLEM DESCRIPTION AND STATISTICAL MODEL

We begin our analysis by considering an SDDS based on the use of CoAP configured to operate with non-confirmed messages, and using UDP as transport protocol. It consists of an *unreliable soft-state* (SS) *SDDS* protocol. Let an *information record* (IR) indicate the atomic set of information, or message, to be distributed (e.g. the CoAP client request) by an *outbound* SDDS. According to a *generic* SS management, upon receiving an IR update from an application, the sender (i.e., the CoAP client) transmits it to receivers unreliably through IR *update request* messages. Any sensor receiving an update message stores the received IR specifying the management policy of the collected information and associates a IR *timeout T* to it. It could represent, for instance, the request to register the sender as an observer at the sensor, so as to be notified automatically when a predefined condition occurs, which is an option of CoAP. The sender refreshes this IR periodically by specific IR *refresh* messages, transmitted unreliably. The IR timer is reset to its initial value if a refresh message is received by its expiration, otherwise the IR is deleted. When the IR is removed from the sender, it stops sending refresh messages thus causing IR removal at all receivers.

A finite state diagram modeling all events is shown in Fig. 1. The state of the model is determined by symbol pair, depending on the IR stored in the sender and in receivers, respectively. The symbol "⊗" denotes a valid stored state, while the symbol "−" denotes either an invalid IR or no IR stored. An IR is considered *valid* only if it is stored in the sender. In case the IR stored in the sender is removed or changed, all IRs already stored in the network become invalid.

Some of the events determining the system evolution, such IR updates, happen in a substantially random fashion. Hence, their inter-arrival time may be modeled as an exponential random variable. Some other events, such as the message propagation through the network, are characterized by nearly constant duration. In case of adjacent nodes, the propagation time is typically very close to its average value. In case of multiple hops staying between two corresponding nodes, the propagation time is substantially variable, but this variability cannot typically be modeled as an exponential memoryless random variable. These events correspond to state transitions indicated by dashed arrows in Fig. 1.

In what follows we describe the model shown in Fig. 1 step by step, starting from the state 1: (−,−).

*State 1:* (−,−). This state represents the absence of a valid IR in the sender, thus even in the network. No messages are being transferred through the network. It is possible to exit this state only if the application generates an IR and stores it in the sender. We assume that the IRs are generated or updated in a totally random fashion and independently of each other. Hence, we model the state 1 holding time as an exponentially distributed random variable with parameter $\lambda_u$, which is the signaling generation/update rate. Upon IR generation or update, a transition to the model state 3, (⊗,−), happens.

*State 3,* (⊗,−). This state indicates that an IR has been generated or updated at the sender. IR update messages are immediately sent to receivers. Three events may cause a state transition. (i) The IR lifetime expires and it is removed from the sender by the application. The inter-arrival time of such events is modeled as an exponential random variable with rate $\lambda_d$. It causes a transition back to the state 1. (ii) One or more IR update messages are lost, hence not all receivers can store a valid IR. This event causes a transition to the state 4, (⊗,−)$_2$. Under the hypothesis of independent transmissions, the probability of this event is 1-(1-$P_{loss}$)$^N$, where $N$ is the number of receivers and $P_{loss}$ is the probability that a message is lost during its way to the receiver. Clearly, we have assumed statistically independent loss events, even if this hypothesis could be not appropriate for some SDDSs.

Nevertheless, without any further information about network configuration we cannot be more precise. Let $D$ denote the maximum end-to-end message transfer time, which includes both transmission and propagation times. Thus, the state transition rate from the state 3(⊗,−) to the state 4 (⊗,−)$_2$ can be written as

$$E_4 = \frac{1-(1-P_{loss})^N}{D} \qquad (1)$$

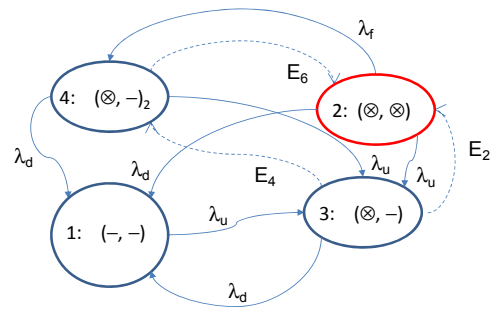

Fig. 1. State diagram of a generic SDDS.

(iii) All IR update messages are successfully received. A transition from the state 3 (⊗,−) to the state 2 (⊗,⊗) happens, characterized by a rate

$$E_2 = \frac{(1-P_{loss})^N}{D} \qquad (2)$$

*State 4* (⊗,−)$_2$. This state indicates that a valid IR is stored at the sender, not all receivers store a valid IR, and no update messages are being transferred. Only IR refresh messages are sent periodically. Three possible events may cause a state transition. (i) The valid IR is removed by the application from the sender. A transition to the state 1 (−,−) happens. As mentioned above, the inter-arrival time of such an event is modeled as an exponential random variable with rate $\lambda_d$. (ii) An IR refresh message is successfully received by all receivers that have previously failed in receiving an IR update message. A state transition from the state 4 (⊗,−)$_2$ to the state 2 (⊗,⊗) happens, characterized by a rate

$$E_6 = \frac{(1-P_{loss})^{NP_{loss}}}{T} \qquad (3)$$

being $T$ the IR refresh period. Note that in equation (3) we have estimated the number of receivers that can cause a state transition after a successful reception of an IR refresh message equal to the expected number of receivers that have not received the IR update message, $N \cdot P_{loss}$. For simplicity we have not considered the event of a receiver that successfully receives an IR refresh message after multiple failings. Since the probability of such an event is a function of powers of $P_{loss}$ greater than one, it is negligible for typical $P_{loss}$ values. (iii) IR generation or update at the sender. As above, we model the inter-arrival time of such events, which cause a transition to the state 3 (⊗,−), as an exponentially distributed random variable with parameter $\lambda_u$.

*State 2* (⊗,⊗). In this state both the sender and all receivers store the same valid IR. Thus, this state corresponds to *consistency*. Three possible events may cause a state transition. (i) An IR removal at the sender by the application when its lifetime expires. As mentioned above, the inter-arrival time of such an event is modeled as an exponential random variable with rate $\lambda_d$. It causes a state transition to the state 1 (−,−).

(ii) An *erroneous* IR removal at receivers, which causes a transition to the state 4 (⊗,−)₂. This event corresponds to the failed reception of all refresh messages by the IR timeout $T$. Since such events should be very sporadic in systems of acceptable quality, their inter-arrival time can be modeled as an exponential random variable with rate $\lambda_f$. Under the assumption of uncorrelated loss events, the value of $\lambda_f$ can be expressed as:

$$\lambda_f = \frac{\prod_{i=0}^{\left\lfloor \frac{X}{T} \right\rfloor} \left(1 - (1 - P_{loss})^{NP_{loss}^i}\right)}{X} \quad (4)$$

(iii) An IR generation/update from the application, modeled as the outcome of an exponentially distributed random variable with parameter $\lambda_u$. It causes a transition to the state 3 (⊗,−).

It is easy to show that a simple change of the semantic of the states shown in Fig. 1 makes the model shown suitable for modeling also an inbound SDDS.

The first objective of this paper is to evaluate the steady state probability distribution of the model, from which the outbound probability of consistency immediately follows, which equals the steady state 2 (⊗,⊗) probability. One may argue that this probability should be conditioned to the presence of a valid IR in the network, which means normalizing it to the state 1 probability. Nevertheless, we prefer avoiding this normalization since it would not highlight scarce network utilization, corresponding to large state 1 probability. Nevertheless, in what follows we will show the probability of all states in order to make the whole information available to the reader.

From the description above it follows that the state holding times cannot be modeled as exponential random variables. A more suitable model consists of a *continuous-time semi-Markov process*, first introduced by Lévy [8] and Smith [9]. According to this model, the state holding times may have a generic distribution, depending on the two states between which the transition is made.

Formally, given a countable set of states $S$, let $S_0, S_1, S_2,\ldots$, be a sequence of visited states, with $S_i \in S$, and let $0=T_0, T_1, T_2,\ldots$, be the times of transitions into each state of the sequence. A Semi-Markov process satisfies the following property:

$$\Pr(S_{n+1} = j, T_{n+1} - T_n \leq t \mid S_0,\ldots,S_n; T_0,\ldots,T_n) = \Pr(S_{n+1} = j, T_{n+1} - T_n \leq t \mid S_n). \quad (5)$$

In addition, the untimed sequence of state transitions $S_0, S_1, S_2,\ldots$ is a realization of a discrete Markov process embedded in the continuous process.

In our model we need to represent nearly constant delays corresponding to nearly constant message transfer times. These delays generate upper bounds to holding times in some states. For example, the state 3 holding time is upper bound by the transfer time $D$, but if the IR lifetime expires before the update message is received, a transition to state 1 happens. Note that in our model all distributions that govern transitions are initialized each time a state is entered. Otherwise, we should have modeled it as a *generalized* semi-Markov process [6].

In order to represent this behavior, we resort to phase-type distributions to model the state sojourn time. In particular, we opted for the Erlang distribution since it has been proved to be the phase-type distribution that minimizes the coefficient of variation $C=(\sigma_x/m_x)^2$ [5], where $\sigma_x$ and $m_x$ are the standard deviation and the mean value of a generic random variable $x$. In particular, the coefficient of variation of an Erlang-k distribution generated by a cascade o $k$ exponential i.i.d. random variables equals $1/k$ [5]. Hence, each nearly constant delay, say $\delta$, has been approximated by the sum of a number $k$ of exponential random variables, each having an average rate of $k/\delta$. The intuition behind this approximation is well illustrated by using the moment generating function of an Erlang-k distribution with rate $k/\delta$, which is

$$X(s) = \left(1 + \frac{\delta}{k}s\right)^{-k} \underset{k \to \infty}{\to} e^{-s\delta} \quad (6)$$

Thus, an Erlang-k distribution has a moment generating function converging towards the moment generating function of a deterministic random variable. The speed of convergence is $O(1/k)$, which is the best achievable by using phase-type distributions.

In the SDDS types analyzed in what follows, our calculations has determined the suitable number of Erlang phases for modeling nearly constant delays. It is the minimum number of phases sufficiently large so as any further increase does not produce significant variations in the computed state probability distribution.

Through this model of the state holding time, the state diagram shown in Fig. 1 can be represented as shown in Fig. 2. In what follows this model will be referred to as model 1.

Transitions to states 3 (⊗,−) and 4 (⊗,−)₂ enter the Erlang states $3_{in}$ and $4_{in}$, respectively. IR updates and expiration events can cause transitions exiting states 3 and 4 at any Erlang state.

Now we analyze the impact that the Erlang-k approximation of the state sojourn time can have on the evaluation of the probability of consistency. We focus on events having an exponential inter-arrival distribution with rate $\lambda$ ($\lambda=\lambda_u$ or $\lambda=\lambda_d$). First, we evaluate the probability of no occurrence of such events during an Erlang state holding time.

Due to memoryless nature of such events, this probability may be evaluated as the product of the probability that no events happen during each exponentially distributed phase periods $\tau_i$, $i=1,\ldots k$, the sum of which forms an Erlang-k outcome. Since the rate of each exponential distribution is

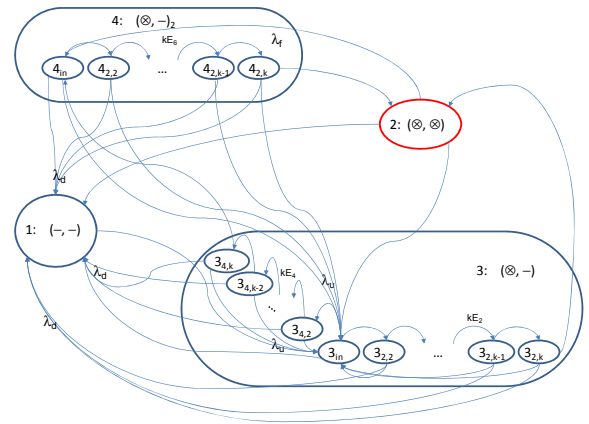

Fig. 2. State diagram of the model 1, an Erlang approximation of the semi-Markov process.

$\mu = k/\delta$, the desired probability is:

$$P_0(k) = \prod_{i=1}^{k} \int_0^\infty \mu e^{-\lambda \tau_i} e^{-\mu \tau_i} d\tau_i = \left(1 - \frac{\lambda \delta}{k}\right)^{-k} \xrightarrow[k \to \infty]{k \to 1} \begin{array}{c} (1-\lambda\delta)^{-1} \\ e^{-\lambda\delta} \end{array} \quad (7)$$

As expected, when $k$ increases the probability $P_0(k)$ converges to the probability that no evens happen during a constant propagation time $\delta$. If k=1, $P_0(1)$ corresponds to the average probability that no events happen during a single exponentially distributed propagation time, i.e. a Markovian approximation of the state sojourn time.

Since $P_0(k)$ is a monotonically decreasing function of k, it follows that the probability of occurrence of an event at rate $\lambda_u$ or $\lambda_d$ when the SDDS is in state 3 ($\otimes$,–) or state 4 ($\otimes$,–)$_2$ increases as k increases. As Fig. 2 shows, this means when k increases the flow of probability both increases towards the sub-state $3_{in}$, thus increasing the overall state 3 holding time, and decreases towards the state 2. Hence, we can conclude that the Erlang approximation of the state sojourn time provides an over-estimate of the probability of consistency, and the simple Markovian approximation provides an upper bound of it. In what follows we evaluate the probability of consistency analytically, and show that results are in accordance with our conclusions. Results will be confirmed by determining the stationary distribution of the states of the continuous-time semi-Markov process represented in Fig. 1 analytically by using the method of the *embedded Markov chain*. Space limitations do not allow us to shows the detailed derivation.

Let **A** denote the *infinitesimal generator*, or *rate matrix*, of the state diagram depicted in Fig. 2. Only non-zero entries are reported. This matrix allows writing the forward *Chapman-Kolmogorov* equations:

$$\frac{d\mathbf{X}(t)}{dt} = \mathbf{X}(t)\mathbf{A} \qquad (8)$$

where **X**(t) is a 1×3k+2 row vector, the components of which are the time-variable probabilities of the states in Fig. 2.

The solution of (8) is $\mathbf{X}(t) = \mathbf{X}_0 e^{\mathbf{A}t}$, where $\mathbf{X}_0$ is the initial state probability distribution.

The computation of the matrix exponential is prone to severe numerical problems and is still an active research area [13]. In many situation the Laplace transform provides a suitable representation [7], since $e^{\mathbf{A}t} = \mathcal{L}^{-1}(s\mathbf{I} - \mathbf{A})^{-1}$, where **I** is the identity matrix. The steady state probability can be found without inverting the Laplace transform by resorting to the well known *final value theorem*:

$$\lim_{t \to \infty} \mathbf{X}(t) = \lim_{s \to 0} s\mathbf{X}(s) \qquad (9)$$

Since the size of **A** is a function of $k$, the matrix inversion above cannot be done by using either numerical or symbolic tools. Thus, direct computation of **X**(s) as a function of $k$ produced a very long symbolic expression, which is not worth to be shown in this document, especially because it does not provide immediate understanding. The relevant numerical results will be shown in the following section.

The analysis shown above is characterized by a significant complexity. Even if the SS protocol used to implement the SDDS is very simple and message transfer is unreliable, handling the resulting parametric rate matrix **A** in all computations is quite complex. This difficulty is expected to increase considerably if we refine the model by additional states or introduce variations to the SS protocol. Thus, any simplification of the approach that does not significantly impair its reliability is appreciable. We have already found that the Markovian approximation of the state holding time provides an upper bound to the probability of consistency. In some instances this approximation is acceptable, as shown by the analysis of some SDDS instances illustrated in what follows. Nevertheless, in some situations it does not. In addition, any over-estimate of the probability of consistency is never attractive since it does not exclude the possibility that the actual probability of consistency is much lower than its upper bound.

For these reasons, we analyze the possibility of simplifying the model 1 by reducing the accuracy in modeling overlapping events. More precisely, we assume that during a message transfer no other events may happen. The resulting state diagram is shown in Fig. 3. In what follows this model will be referred to as model 2. The limiting state probability distribution can be found as in the previous model, with a simplified rate matrix.

Even if in some situation this approximation is not numerically significant, in particular for short message transfer delay, it has an important conceptual consequence. This model cannot converge to the real SDDS operation for any value of the number $k$ of the Erlang phases. In fact, the probability of no events having an exponential inter-arrival distribution happening during the message transfer is

$$\hat{P}_0(k) = \int_0^\infty \mu e^{-\lambda \tau_i} e^{-\mu \tau_i} d\tau_i = \left(1 - \frac{\lambda D}{k}\right)^{-1} \xrightarrow[k \to \infty]{k \to 1} \begin{array}{c} (1 - \lambda D)^{-1} \\ 1 \end{array} \quad (10)$$

Thus, for $k$=1 also this model reduces to the mere Markov approximation. Differently, For increasing $k$ values the sum of the probabilities of the state 3 and 4 increases. Since these states affects the flow probability entering the state 2, we conclude that the estimated probability of consistency decreases when $k$ increases.

### III. CASE STUDIES AND RESULTS

In order to assess the reliability of the models proposed it is useful to identify some case studies which differ by some key parameter values. Space limitations do not allow presenting a lot of results. Nevertheless, Table I reports two sets of configuration values suitable to give appropriate insights.

Fig. 4 shows the steady state probability of the SDDS configurations. Solid lines are relevant the model 1 depicted in Fig. 2. Dashed lines are obtained by using the simplified model 2, shown in Fig. 3. As mentioned above, we identify the probability of consistency with the state 2 probability. As expected, solid and dashed curves depart from the same value for $k$=1, which corresponds to the Markov approximation of the holding time. Results are also in accordance with the conclusions that the consistency probability is a monotonically decreasing function of $k$ and the Markovian approximation provides an upper bound of it. Due to the converging properties of the model 1, demonstrated above, we regard the converging value of state 2 solid curve as the most reliable value of the probability of consistence. This conclusion is confirmed by the probability of consistency determined analytically by using the method of the embedded Markov chain, shown by an arrow in Fig. 4 and in subsequent graphs. Thus, we will use it as the reference value for further considerations.

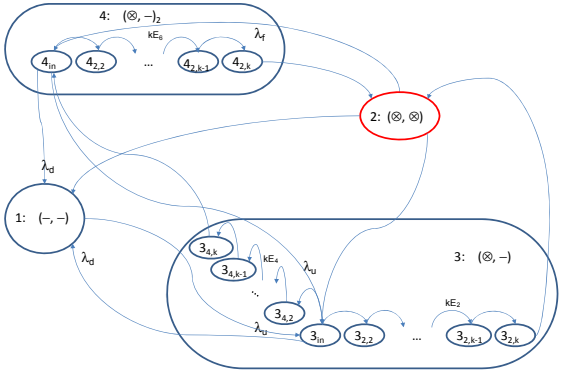

Fig. 3. State diagram of the simplified Erlang approximation.

Even if we are essentially interested to the state 2 probability, all figures shown below include also the probability of the other states. This information is useful to argue the preferred direction of the probability flow when the estimated state 2 probability diverges from the reference value. When the message transfer delay is very short (SDDS Case 1), the Markov upper bound is a very good approximation of the reference value. This is intuitive, since the standard deviation of the exponential distribution is equal to its mean, and when the latter decreases the approximation improves accordingly. When the probability of consistency decreases, the upper bound provided by the Markov approximation is significantly larger than the reference value. Thus, in critical situations, the use of the Markovian approximation is misleading. The intuition behind this behavior is that the factors causing a deterioration of the SDDS consistency tend to steer the probability flows towards states 3 and 4. Being indeed the holding time of these two states those affected by the Markovian approximation, its effects are made evident. In this situations the approximation of the simplified Erlang model 2 is preferable than the Markovian one, since the monotonically decreasing shape of the curves decrease the probability of over-estimating the SDDS consistency. Nevertheless, even if in some conditions the estimates provided by the simplified Erlang model 2 are acceptable, in other configurations this model could significantly under-estimate the probability of consistency. Even if an under-estimation is not as detrimental as an over-estimation, it could result not negligible.

Message transfer reliability can be introduced by different solutions implemented at different protocol layers. They could be byte-oriented, packet-oriented or message-oriented. Even retransmissions can follow different rules. Modeling all these solutions is hard and somewhat useless for the sake of this general analysis. In fact, their differences may be due to other reasons, such as implementation issues, or efficiency in the use of network resources, and not strictly determined by consistency reasons. Without loss of generality, we assume to implement reliability in the CoAP protocol, by using confirmed messages over UDP. Thus, a message is retransmitted if a positive acknowledgement is not received by a round trip time. In addition, we assume that refresh messages are no longer used, since we introduced reliability. The implications in our models consists of modifying the transition rate from state 4 to state 2, which becomes:

TABLE I. SDDS CONFIGURATION VALUES

| SDDS type | $\lambda_u$ | $\lambda_d$ | $\lambda_f$ | $P_{loss}$ | #nodes | D(s) | T(s) |
|---|---|---|---|---|---|---|---|
| Case 1 | 1 | $5 \cdot 10^{-3}$ | $2 \cdot 10^{-8}$ | $10^{-3}$ | 100 | 0.01 | 5 |
| Case 2 | 0.1 | $5 \cdot 10^{-3}$ | $2 \cdot 10^{-8}$ | $10^{-3}$ | 100 | 1 | 10 |

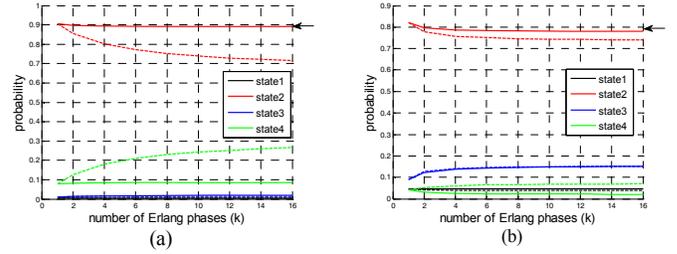

Fig. 4. Probability of consistency vs. number of Erlang Phases. Case studies 1 (a) and 2 (b). Solid and dashed lines are obtained by using the model 1 and the model 2, respectively. Unreliable transmission. Arrows indicate the probability of consistency computed analytically.

$$E_{6,rel} = \frac{(1-P_e)^{NP_{loss}}}{2D} \qquad (11)$$

We have analyzed once again all SDDS types reported in Table I. The results achieved by this change are depicted in Fig. 5. The predictable result is that the SDDS consistency has improved. A further relevant result is that in all situations the approximations of the simplified Erlang model are much better than those obtained by using the Markovian model. In most of the situations analyzed the approximation of the simplified Erlang model is actually negligible. The intuition behind this behavior is that the increased rate of transition from the state 4 to the state 2 has decreased the importance of modeling the transitions from all the sub-states of the state 4 to the sub-state $3_{in}$ and to the state 1. In addition, the exponentially distributed transitions with the highest rate ($\lambda_u$), departing from sub-states in the state 3, are directed to the sub-state $3_{in}$, i.e. within the same state, and their presence do not alter the total state 3 probability significantly. No other states are affected by the simplified Erlang model. Thus, this model provides estimates very close to the complex model 1, although a considerable reduction of the complexity of the model.

All considerations done so far are relevant to steady state probabilities. It is legitimate to wonder about the dynamic behavior of the systems. Space limitations do not allow including a detailed analysis. Nevertheless, Fig. 6 shows the evolution of the state probability distribution of the SDDS Case 2, unreliable. The probability distribution has been computed through the simplified Erlang model with $k=10$. Since the $k$ value is fixed, it was easy to determine the matrix exponential numerically through the well designed *expm( )* Matlab function. The starting time is the time of generation of the first IR, i.e. the initial state in the state 3. First, it is worth to note that the state probability distribution converges to the one shown in Fig. 4 for $k=10$, which is a further validation of the results achieved. Then, we observe that the consistency probability becomes close to the reference value in very few seconds.

The importance of these results goes beyond the scope of this paper. We can observe that in particular situations the reliable message transfer is not enough to guarantee a satisfactory probability of consistence. In such situations additional techniques must be introduced for improving the

SDDS reliability, both at architectural and at protocol levels. In order to analyze these solutions, further states must be introduced in the model, with generic holding time. Having deeply understood the mechanism that make the simplified Erlang model reliable, it is a suitable candidate for analyzing more complicate SDDSs with a manageable complexity.

## IV. RELATED WORK

Consistency analysis is an active research topic in different areas. For example, the cache consistency problem for ad-hoc and wireless networks has been faced in [10].

A middleware solution for developing replicated systems with various consistency requirements is described in [11].

The use of the consistency metric for analyzing soft state protocols has initially been proposed in [2], where the authors presented a model based on Jackson queuing networks. A contribution to this research was also given in [1], where the authors compare a hard-state and a number of soft state protocols with different levels of reliability. Our method of analysis of the SDDS consistency probability was inspired by the state model shown in [1] for the single-hop SS analysis. Our approach differs from [1] in the state semantic, the equivalence of invalid states and absence of states, and in the model of nearly constant state holding time. In [1] it has been modeled by Markovian distribution, and the impact of the approximation has been evaluated by simulations. We have preferred to make use to the converging properties of the Erlang distribution.

For what concerns related statistical models, a comprehensive analysis of approximations of semi-Markov distribution through phase-type distributions is shown in [14]. A useful analysis highlighting long run average properties of generalized semi-Markov processes with both fixed-delay and variable-delay can be found in [6].

## V. CONCLUSIONS.

In this paper we have shown and analyzed a general framework for analyzing the consistency of Sensor Data Distribution Systems. We have proposed different approaches for modeling general state holding times. The simplest one is a mere Markov approximation. Then, we have proposed an Erlang phase-type approximation of the state holding time with exponentially distributed transitions at each Erlang sub-states. We have shown analytically the convergence properties of this model to deterministic state holding time. Given the complexity of this model, we have proposed and analyzed also a simplified Erlang model.

The results of our analysis is that for very short message transfer time, the simple Markov approximation provides good estimates of the probability of consistency. When the message transfer time increases, the Markov approximation tends to significantly overestimate of this probability. In this case the Erlang-based model is preferable, and the simplified version is acceptable. When messages are transferred through reliable techniques, the simplified Erlang model provides very good estimates, that make it a suitable approach for supporting future research directions, applying them also to environments other than SDDS, such as general signaling frameworks [12][16].

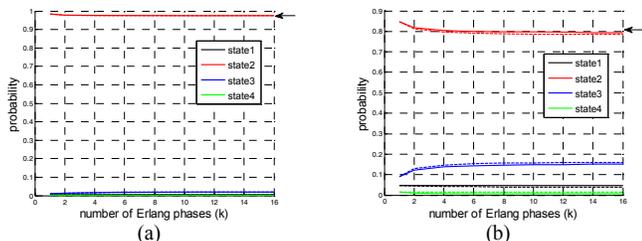

Fig. 5. Probability of consistency vs. number of Erlang Phases. Case studies 1 (a) and 2 (b). Solid and dashed lines are obtained by using the model 1 and the model 2, respectively. Reliable transmission. Arrows indicate the probability of consistency computed analytically.

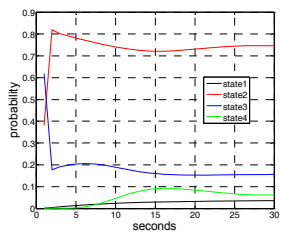

Fig. 6. State probability vs. time. Case study 2. Estimates obtained by using the simplified model 2. Unreliable transmission.